\documentclass[aps,twocolumn,superscriptaddress,showpacs,letter]{revtex4-1}
\usepackage[utf8]{inputenc}
\usepackage{graphicx}
\usepackage{dcolumn}
\usepackage{bm}
\usepackage{cancel}
\usepackage{color}             
\usepackage{epsfig}
\usepackage{amssymb}
\usepackage{amsmath}
\usepackage{multirow}
\usepackage{hyperref}
\usepackage{cleveref}

\begin{document}

\title{Soft Pomeron in light of the LHC correlated data}

\author{M.~Broilo}
\email{mateus.broilo90@gmail.com}
\affiliation{Instituto de F\'isica e Matem\'atica, Universidade Federal de Pelotas, \\ 96010-900, Pelotas, Rio Grande do Sul, Brazil}
\author{D.~A.~Fagundes}
\email{daniel.fagundes@ufsc.br}
\affiliation{Departamento de Ci\^encias Exatas e Educa\c{c}\~ao, Universidade Federal de Santa Catarina - Campus Blumenau, \\ 89065-300, 
Blumenau, Santa Catarina, Brazil}
\author{E.~G.~S.~Luna}
\email{luna@if.ufrgs.br}
\affiliation{Instituto de F\'isica, Universidade Federal do Rio Grande do Sul, Caixa Postal 15051, 91501-970, Porto Alegre, Rio Grande do Sul, Brazil}
\author{M.~Pel\'aez}
\email{mpelaez@fing.edu.uy}
\affiliation{Instituto de F\'{\i}sica, Facultad de Ingenier\'{\i}a, Universidad de la Rep\'ublica, \\
J.H. y Reissig 565, 11000 Montevideo, Uruguay}
 

\begin{abstract}

The LHC has released precise measurements of elastic proton-proton scattering that provide a unique constraint on the asymptotic behavior 
of the scattering amplitude at high energies. Recent reanalyses of part of these data indicate that the central values of some forward
quantities would be different than initially observed. We introduce correlation information between the original and the reanalyzed data
sets in a way suitable for a global fitting analysis of all data. 
The careful treatment of correlated errors leads to much less stringent limits on the 
$\rho$ uncertainty and sets up the stage for describing the forward data using a scattering amplitude dominated
by only crossing-even terms.
In the light of these correlated data we determine the parameters of the soft Pomeron from the Regge theory. We use Born-level and
eikonalized amplitudes. In the Born-level case we estimate the contribution of the
double Pomeron exchange, while in the latter case we investigate the role of the eikonalization in both the one- and two-channel models.
The role of the proton-Pomeron vertex form and of the
nearest $t$-channel singularity in the Pomeron trajectory receives particular attention. We discuss the implications of our results and
present predictions for the total cross section and the $\rho$ parameter in
proton-proton collisions at LHC and cosmic ray energies.

\end{abstract}


\maketitle

\section{Introduction}

Diffractive processes account for a substantial fraction of hadron-hadron total cross sections at high energies. These processes, which
include elastic scattering or single- or double-diffractive dissociation, are characterized by the presence of one or more large
rapidity gaps, which in turn were, for many years, usually associated with the exchange of a colorless state having the quantum numbers
of the vacuum:
the Pomeron ($\Bbb P$). Recently, the TOTEM experiment at the LHC released measurements at $\sqrt{s} = 13$ TeV of the ratio
real to imaginary parts of the forward scattering amplitude, namely, $\rho = 0.09\pm 0.01$ and $\rho = 0.10\pm 0.01$ \cite{TOTEM008}. These
values, as compared with measurements at lower energies and predictions from a wide variety of phenomenological models, suggest that a
crossing-odd elastic term may play a central role in soft interactions at high energies.
Since then, an intense debate has centered around the asymptotic nature of the $C$-parity of the scattering amplitude. 
Specifically,
the TOTEM results have triggered an extensive discussion on the question of whether or not the combined behavior of $\sigma_{tot}$ and
$\rho$ at high energies is a manifestation of the so-called {\it Odderon} \cite{odderon002}, the $C=-1$ partner of the $C=+1$ Pomeron
\cite{odderon001}. In the QCD language, the Odderon can be associated to a colorless $C$-odd $t$-channel state with an intercept at or
near one \cite{genia001,contreras001}.

However, recent reanalyses of the TOTEM results for the Coulomb-nuclear interference (CNI) region have shown that the values
of $\rho$ at $\sqrt{s} = 8$ and 13 TeV may be larger than those reported by the TOTEM Collaboration \cite{ezhelapetr01,panchsriva01}.
In one of these reanalyses \cite{ezhelapetr01}, using the same nuclear amplitude used by TOTEM but a modified formula for the CNI term,
the value $\rho = 0.123\pm 0.010$ is obtained at $\sqrt{s} = 13$ TeV; 
in the other \cite{panchsriva01}, by means of a modified version of the nuclear amplitude
\cite{sriva01}, the values $\rho = 0.135$ and $\rho = 0.137$ are obtained at $\sqrt{s} = 8$ TeV, whereas the values $\rho = 0.133$ and
$\rho = 0.134$ are obtained at $\sqrt{s} = 13$ TeV. We remark at this point that 
in a global fitting analysis we cannot simply choose any of these results over another, nor choose all the results while neglecting the
systematic correlation between them. We must also remember that correlation information is already necessary in dealing with part of the
original TOTEM data. For example, the measurements of $\rho$ at $\sqrt{s}=13$ TeV were
extracted {\it from the same set} of differential cross section
data and have their central values depending on different physics assumptions. In this paper, we show that it is possible to include
correlated systematics and combine all these data in a meaningful way. For this, we apply the standard procedure adopted by
the Particle Data Group in treating correlated errors \cite{PDG}. As we will see, after the inclusion of the correlated systematics,
each measurement may be treated as independent and, as a result, averaged as usual with other data.

In this work, we are also interested in the effects of the unitarization on the soft Pomeron parameters. Then after fitting analyses
using only Born-level scattering amplitudes, we investigate the consequences of including eikonalization in one-channel models. Finally,
we repeat our eikonal analyses taking into account only two-channel amplitudes. It is worth remarking on the fact that,
in general, only one of these three approaches is usually adopted as a default, that is, some authors focus specifically on
Born-level calculations (by arguing that unitarity violation only occurs far above the LHC energies), while others focus solely on
one-channel eikonal amplitudes. The use of different procedures made results difficult to compare. Hence, in this work, we adopt all these
approaches, making not only the comparison of the results more transparent, but also allowing us to follow the evolution of the Pomeron
parameters during the transition from Born-level to eikonalized amplitudes.

In summary, we have shown that the LHC data, after the introduction of correlation information, provide a unique constraint on the Pomeron
parameters and allow us to study its behavior more thoroughly since the contribution of the Pomeron component to $\chi^{2}$ is
absolutely dominant in the LHC regime.
Moreover, the small value of $\alpha^{\prime}_{\Bbb P}$ usually obtained from screened Regge models
indicates that the soft Pomeron may be treated perturbatively, since in the Gribov Reggeon calculus the mean transverse momentum of the
partons is given by $\langle p_{T} \rangle = 1/\sqrt{\alpha^{\prime}_{\Bbb P}}$ \cite{gribov001,baker001}. This perturbative approach 
raises the possibility for building a fundamental theory for soft processes based upon QCD. 
Additionally, screening effects can be
calculated in terms of a
two-channel eikonal model and again the correlated LHC data are instrumental in determining the effects of the eikonalization on the
Pomeron
parameters in both the one- and two-channel models. Thus, given the central role that the soft Pomeron plays in strong processes, its
close scrutiny continues to be a core task in hadron physics. 

The outline of this paper is as follows. In Sec. II, we introduce the Regge formalism used to model Born-level amplitudes. Within this
approach, we investigate the single as well as the double Pomeron exchange. In Sec. III, we investigate the role of the eikonalization
procedure in Regge amplitudes. We study both one- and two-channel eikonal models. In Sec. IV, we discuss the PDG procedure for
treating correlated errors and present our results. In the last section, we draw our results and conclusions.

\section{Born-level Amplitudes}

In the soft regime, i.e. small $t$ domain, diffractive processes are described by Regge theory \cite{barone001}, in which the
high-energy behavior of the scattering amplitude is described by singularities of the amplitude in the complex plane of angular momentum
$j$. 
In the simplest scenario, the diffractive processes are driven by an isolated pole
at $j=\alpha(t)$, resulting in
an elastic amplitude ${\cal A}(s,t)$ written in terms of the Regge pole trajectory $\alpha(t)$, namely,
${\cal A}(s,t) \propto s^{\alpha(t)}$. If more than one pole contributes, the elastic scattering amplitude is expressed in the
$s$-channel as a descending asymptotic series of powers of $s$,
\begin{eqnarray}
{\cal A}(s,t)=\sum_{i}\gamma_{i}(t)\eta_{i}(t)s^{\alpha_{i}(t)} ,
\label{equation01}
\end{eqnarray}
where $\gamma_{i}(t)$ is the residue function and $\eta_{i}(t)$ is the signature factor. Each term in (\ref{equation01}) represents a
specific exchange in the $t$-channel. From the optical theorem, the total cross section reads
\begin{eqnarray}
\sigma_{tot}(s)=\sum_{i}4\pi g_{i} s^{\alpha_{i}(0)-1} ,
\label{equation02}
\end{eqnarray}
where $g_{i} \equiv \gamma_{i}(0)\textnormal{Im}\{ \eta_{i}(0)\}$. The Pomeron as it emerges from fits to forward observables is
called {\it soft Pomeron}. The magnitude of its intercept plays a central
role in Regge theory, since the Pomeron is the pole with the largest intercept, originally $\alpha_{\Bbb P}(0)=1$. However, in order to
describe the observed increase of all hadronic total cross sections with $s$, the Pomeron should have an effective
intercept such that $\alpha_{\Bbb P}(0)=1+\epsilon$ with $\epsilon > 0$. This {\it supercritical} intercept value is arrived at by taking
into account, in addition to Regge poles, multi-Pomeron cuts in the $j$-plane.

It is well known that good descriptions of forward data up to the Tevatron energy have been obtained by using a linear Pomeron trajectory
\cite{dl001,cudell001,goulianos001,luna001,cudell003}, namely, $\alpha_{\Bbb P}(t) = 1+\epsilon +\alpha^{\prime}_{\Bbb P}t$.
The energy dependence of the total and diffractive cross sections is driven by $\epsilon$ while $\alpha^{\prime}_{\Bbb P}$ determines the
energy dependence of the forward slopes. We must note, however, that ZEUS and H1 small-$t$ data for exclusive $\rho$ and $\phi$
photoproduction call forth a rather nonlinear Pomeron trajectory \cite{zeush1001}.
We shall see that the data with correlated systematics 
allow us to address more effectively the question of linearity versus nonlinearity of the Pomeron trajectory.

The forward Born-level Regge amplitude introduced some time ago by Donnachie and Landshoff has two contributions \cite{dl001}, one
representing an effective single Pomeron and the other representing the exchange of the highest-spin meson trajectories
($a_{2}$, $f_{2}$, $\omega$ and $\rho$). However, more recent analysis has indicated that the assumption of degeneracy of the
mesons trajectories is not supported by the forward data \cite{cudell001,goulianos001,luna001}. The best results are obtained with
a Born-level amplitude decomposed into three contributions,
\begin{eqnarray}
{\cal A}_{Born}(s,t) = {\cal A}_{\Bbb P}(s,t) + {\cal A}_{+}(s,t) + \tau {\cal A}_{-}(s,t) ,
\label{amplit01}
\end{eqnarray}
where $\tau$ flips sign when going from $pp$ ($\tau=-1$) to $\bar{p}p$ ($\tau=+1$). The term ${\cal A}_{\Bbb P}(s,t)$ represents the 
exchange of the Pomeron, ${\cal A}_{+}(s,t)$ the exchange of the Reggeons with $C=+1$ ($a_{2}$ and $f_{2}$), and ${\cal A}_{-}(s,t)$ that
of the Reggeons with $C=-1$ ($\omega$ and $\rho$). Specifically, the amplitude for single exchange is
\begin{eqnarray}
{\cal A}_{i}(s,t) = \beta_{i}^{2}(t)\eta_{i}(t)\left( \frac{s}{s_{0}} \right)^{\alpha_{i}(t)} ,
\label{equation04}
\end{eqnarray}
$i=\Bbb P, +, -$, where $\beta_{i}(t)$ is the elastic proton-Reggeon vertex, $\eta_{i}(t)$ is the signature factor, and
$\alpha_{i}(t)$ is the Regge pole trajectory. Here $s_{0}$ is a mass scale usually chosen to be $s_{0} = 1$ GeV$^{2}$.
By comparing the Eqs. (\ref{equation01}) and (\ref{equation04}), it can be seen that
the residue function factorizes as $\gamma_{i}(t)=\beta_{i}^{2}(t)$.
The signature factor, which completely defines the phase of the scattering amplitude, is given by \cite{barone001}
\begin{eqnarray}
\eta_{i} (t) = - \frac{1+\xi \, e^{-i\pi \alpha_{i}(t)}}{\sin (\pi \alpha_{i}(t))} ,
\end{eqnarray}
where $\xi=+1$ for the Pomeron and the Reggeons $a_{2}$ and $f_{2}$, and $\xi=-1$ for the Reggeons $\omega$ and $\rho$. Thus, the $pp$
and $\bar{p}p$ scatterings are described in terms of Pomeron, positive and negative signature Regge exchange. However, in order to
simplify the numerical calculations involved in the forthcoming eikonal analyses, we adopt in this work
$\eta_{i}(t)=-e^{-i\frac{\pi}{2}\alpha_{i}(t)}$ for
even-signature trajectories and $\eta_{i}(t)=ie^{-i\frac{\pi}{2}\alpha_{i}(t)}$ for odd-signature ones \cite{goulianos001}. The choice of
these simplified signatures does not affect our results since the numerical integrals are strongly dominated by the region where $t$
is very small.

The positive-signature secondary Reggeons ($a_{2}$ and $f_{2}$) are taken to have an exponential form for the proton-Reggeon vertex, 
\begin{eqnarray}
\beta_{+}(t)=\beta_{+}(0)\exp (r_{+}t/2 ) ,
\end{eqnarray}
and to lie on an exchange-degenerate linear trajectory of form 
\begin{eqnarray}
\alpha_{+}(t) = 1 - \eta_{+} + \alpha^{\prime}_{+} t.
\end{eqnarray}
Similarly, the negative-signature secondary Reggeons ($\omega$ and $\rho$) are described by the pa\-ra\-me\-ters
$\beta_{-}(0)$, $r_{-}$, $\eta_{-}$, and $\alpha^{\prime}_{-}$. 

For Pomeron exchange, we investigate two different types of proton-Pomeron vertex and two different types of trajectory, one of which being nonlinear. Our philosophy is, using the standard statistical $\chi^{2}$ test, to evaluate the
relative plausibility of these vertices and trajectories in the light of the LHC data, i.e. to consider different combinations of
$\beta_{\Bbb P}(t)$ and $\alpha_{\Bbb P}(t)$, and the effectiveness of these combinations at describing the high-energy forward data. In
the first combination, referred to as ``BI model,'' we adopt an exponential form for the proton-Pomeron vertex,
\begin{eqnarray}
\beta_{\Bbb P}(t)=\beta_{\Bbb P}(0)\exp (r_{\Bbb P}t/2 ) ,
\label{vertex01}
\end{eqnarray}
and a linear Pomeron trajectory,
\begin{eqnarray}
\alpha_{\Bbb P}(t) = \alpha_{\Bbb P}(0) + \alpha^{\prime}_{\Bbb P} t ,
\end{eqnarray}
where henceforth we define $\alpha_{\Bbb P}(0)\equiv 1 + \epsilon$. In the second model, called ``BII,'' we adopt the exponential vertex
(\ref{vertex01}) and the nonlinear Pomeron trajectory \cite{anselm001,kmr001,kmr002,kmr003}
\begin{eqnarray}
\alpha_{\Bbb P}(t) = \alpha_{\Bbb P}(0) + \alpha^{\prime}_{\Bbb P} t + \frac{m_{\pi}^{2}}{32\pi^{3}}\,
h(\tau) ,
\label{pomnlin}
\end{eqnarray}
where
\begin{eqnarray}
h (\tau) &=& -\frac{4}{\tau}\, F_{\pi}^{2}(t) \left[  2\tau - (1+\tau)^{3/2} \ln \left( \frac{\sqrt{1+\tau}+1}{\sqrt{1+\tau}-1} \right)
\right. \nonumber \\
& & + \left. \ln \left( \frac{m^{2}}{m_{\pi}^{2}} \right) \right]  ,
\end{eqnarray}
with $\tau = 4m_{\pi}^{2}/|t|$, $m_{\pi}=139.6$ MeV and $m=1$ GeV. The nonlinear term in the Pomeron trajectory comes from the nearest
$t$-channel singularity (a two-pion loop) \cite{anselm001}. In the above expression, $F_{\pi}(t)$ is the form factor of the pion-Pomeron
vertex, for which we take the standard pole expression $F_{\pi}(t)=\beta_{\pi}/(1-t/a_{1})$. The coefficient $\beta_{\pi}$ specifies the
value of the pion-Pomeron coupling and for this we adopt the additive quark model relation $\beta_{\pi}/\beta_{I\!\!P}(0)=2/3$. In the
third combination, called ``BIII model,'' we adopt the nonlinear Pomeron trajectory (\ref{pomnlin}) and the powerlike form for the
proton-Pomeron vertex
\cite{kmr001,kmr002,kmr003,kmr004}
\begin{eqnarray}
\beta_{\Bbb P}(t)=\frac{\beta_{\Bbb P}(0)}{(1-t/a_{1})(1-t/a_{2})} ,
\end{eqnarray}
where the free parameter $a_{1}$ is the same as the one in the form factor of the pion-Pomeron vertex $F_{\pi}(t)$.
The total cross section, the elastic differential cross section, and the $\rho$ parameter are expressed in terms of the
amplitude (\ref{amplit01}),
\begin{eqnarray}
\sigma_{tot}(s)=\frac{4\pi}{s}\, \textnormal{Im}\, {\cal A}(s,t=0) ,
\label{ertion001}
\end{eqnarray}
\begin{eqnarray}
\frac{d\sigma}{dt}(s,t)=\frac{\pi}{s^{2}}\, \left| {\cal A}(s,t) \right|^{2} ,
\label{ertion002}
\end{eqnarray}
\begin{eqnarray}
\rho(s)=\frac{\textnormal{Re}\, {\cal A}(s,t=0)}{\textnormal{Im}\, {\cal A}(s,t=0)} ,
\label{ertion003}
\end{eqnarray}
where ${\cal A}(s,t) = {\cal A}_{Born}(s,t)$.

It is important to notice that the Pomeron intercept $\alpha_{\Bbb P}(0)=1 + \epsilon$ is an effective power, valid over a limited
range of energies; otherwise, the forward amplitude $A(s,t=0)$ would grow so large that unitarity bound would be violated. Thus, 
the parameter $\epsilon$ represents not only the exchange of a single Pomeron but also $n$-Pomeron exchange processes, $n\geq 2$ \cite{barone001,sergio001}.
These multiple exchanges must tame the rise of $\sigma_{tot}(s)$ so that the breakdown of unitarity is avoided and, as a consequence,
the value of $\epsilon$ should decrease slowly with increasing $s$. The search for a hint of
unitarization breaking up to LHC energies can be verified by investigating the role of multiple Pomeron exchanges on the scattering
amplitude.
Unfortunately, despite the advances in theoretical understanding of the Pomeron in the last four decades, we still do not know how to
do it. Nevertheless, there is a consensus that the contribution of the double Pomeron exchange ($\Bbb P\Bbb P$) is negative and has energy
dependence 
$s^{\alpha_{\Bbb P\Bbb P}(t)}$ divided by some function of $\ln s$ \cite{landshoff001}, where
\begin{eqnarray}
\alpha_{\Bbb P\Bbb P}(t) = 1 + 2\epsilon + \frac{1}{2} \alpha^{\prime}_{\Bbb P}t .
\label{equationdpom02}
\end{eqnarray}
Thus, the $\Bbb P \Bbb P$ contribution is flatter in $t$ than the single $\Bbb P$ exchange, becoming more important for higher
values of $t$. In order to estimate an upper bound on the ratio $R \equiv \beta^{2}_{\Bbb P\Bbb P}(0)/\beta^{2}_{\Bbb P}(0)$,
we add the phenomenological term
\begin{eqnarray}
{\cal A}_{\Bbb P\Bbb P}(s,t) = -\beta_{\Bbb P\Bbb P}^{2}(t)\eta_{\Bbb P\Bbb P}(t)\left( \frac{s}{s_{0}}
\right)^{\alpha_{\Bbb P\Bbb P}(t)} \left[ \ln \left( \frac{-is}{s_{0}} \right) \right]^{-1} \nonumber \\
\label{equationdpom01}
\end{eqnarray}
to the amplitude (\ref{amplit01}), where $\eta_{\Bbb P\Bbb P}(t) = -e^{-i\frac{\pi}{2}\alpha_{\Bbb P\Bbb P}(t)}$ and
$\beta_{\Bbb P\Bbb P}(t) = \beta_{\Bbb P\Bbb P}(0)\exp (r_{\Bbb P}t/4)$. We include this double-Pomeron exchange term
in the model BI. This combination is henceforth called BI+$\Bbb P \Bbb P$ model.

\section{Eikonalized Amplitudes}

\subsection{One-channel amplitudes}

As it was mentioned in the previous section, in the case of Born-level amplitudes, the breakdown of unitarity can be avoided by
introducing the exchange series $\Bbb P + \Bbb P\Bbb P + \Bbb P \Bbb P\Bbb P + \ldots $.
Although some general analytic properties of these multiple-exchange terms are known, it is less clear how to carry out a full
computation of them. On the other hand, it is well established that eikonalization is an effective procedure to take into account
some properties of high-energy $s$-channel unitarity.
In practice, the unitarity of the matrix \textsf{S} in impact parameter ($b$) representation implies the relation \cite{barone001}
\begin{eqnarray}
\textnormal{Im}\, h(s,b) = |h(s,b)|^{2} + G_{inel}(s,b) ,
\label{unitarity001}
\end{eqnarray}
where $h(s,b)$ is the elastic profile function and $G_{inel}(s,b)$, known as the {\it inelastic overlap
function} or {\it shadow profile function}, is the contribution from all inelastic channels. The profile function $h(s,b)$
is related to the elastic scattering amplitude ${\cal A}(s,t)$ by
\begin{eqnarray}
{\cal A}(s,t) = \int_{0}^{\infty}b\, db\, J_{0}(b\sqrt{-t})\, h(s,b) .
\label{scatter001}
\end{eqnarray}
In this picture, we can think of the sum over all inelastic channels as forming a shadow, which ``generates'' elastic scattering.
The unitarity relation
(\ref{unitarity001}) imposes a limit on the elastic profile function, namely, $0\leq |h(s,b)|^{2} \leq \textnormal{Im}\, h(s,b) \leq 1$,
while eikonalization enforces the so-called {\it black-disc} limit: $\textnormal{Im}\, h(s,b) \leq 1/2$. 
The upper value $1/2$ is due to the requirement of a maximal absorption within the eikonal unitarization, in which $h(s,b)$ is written
as
\begin{eqnarray}
h(s,b) = \frac{i}{2} \left[ 1 - e^{i\chi (s,b)}  \right] ,
\end{eqnarray}
with the eikonal function $\chi (s,b)=i\Omega(s,b)/2$ being a purely imaginary function in the limit $s \to\infty$. In other words, at
high energies, the inelastic contribution, $G_{inel}$, dominates and the scattering amplitude ${\cal A}(s,t)$ is predominantly imaginary.
In this regime, $\Omega(s,b) \gg 1$ and $\textnormal{Im}\, h(s,b) = 1/2$. The eikonalization scheme prevents the Froissart-Martin bound
for $\sigma_{tot}(s)$ from being violated. The bound follows from the theorem which states that $\sigma_{tot}(s) \leq C \ln^{2}s$, as
$s\to \infty$, where $C$ is a constant \cite{froissart001}.
The Froissart-Martin bound imposes a strict restriction on the rate of growth of any total cross section.
It is worth mentioning that while the Froissart-Martin bound holds for all eikonalized amplitudes studied in this paper, it is not
necessarily synonymous with {\it total} unitarization: it was shown some time ago that any model for input Pomeron with intercept
$\alpha_{\Bbb P}(0) > 1$ but with linear trajectory is affected by small asymptotic violations of unitarity \cite{giffon001}. We also
notice that eikonal unitarization corresponds to one of the two solutions of the unitarity equation
\begin{eqnarray}
h(s,b) = \frac{1}{2} \left[ 1 \pm \sqrt{1-4G_{inel}(s,b)} \right] ,
\label{unitar008}
\end{eqnarray}
the one with minus sign. Choosing the plus sign in (\ref{unitar008}), we get the alternative solution \cite{tro001}
\begin{eqnarray}
h(s,b) = \frac{\textnormal{Im}\,\tilde{\chi}(s,b)}{1-i\tilde{\chi}(s,b)} ,
\label{unitar009}
\end{eqnarray}
where $\tilde{\chi}(s,b)$ is the analogue of the eikonal $\chi(s,b)$. In this approach, $h(s,b)$ may exceed the black disc limit.
Thus, we see that different unitarization procedures are possible in $b$ representation. In this paper, we follow with the
eikonalization procedure:
the eikonal function is related to the Born amplitude (\ref{amplit01}) by the Fourier-Bessel transform
\begin{eqnarray}
\chi (s,b) = \frac{1}{s}\int^{\infty}_{0} q\, dq\, J_{0}(bq)\, {\cal A}_{Born} (s,t) ,
\label{eikonal007}
\end{eqnarray}
where $t=-q^{2}$; its inverse transform leads to the eikonalized amplitude in $(s,t)$-space,
\begin{eqnarray}
{\cal A}_{eik}(s,t) = is \int^{\infty}_{0} b\, db\, J_{0}(bq) \left[ 1 - e^{i\chi (s,b)} \right] ,
\end{eqnarray}
to be used in the calculation of the observables. Hence, the total cross section, the elastic differential cross section, and the
$\rho$ parameter are calculated using Eqs. (\ref{ertion001})-(\ref{ertion003}) with
${\cal A}(s,t)={\cal A}_{eik}(s,t)$.
In the calculation of the eikonal function (\ref{eikonal007}), the input amplitudes (${\cal A}_{Born} (s,t)$) are simply the ones related
to BI, BII, and BIII models. These one-channel eikonal models are referred to, respectively, as OI, OII, and OIII models.

\subsection{Two-channel amplitudes}

As it was mentioned in Sec. I, an effective Pomeron intercept $\alpha_{\Bbb P}(0) > 1$ is obtained taking into account
multi-Pomeron cuts (moving branch points) in the $j$-plane. These singularities are required in order to assure $s$-channel unitarity.
In the models considered in the preceding text, we have not accounted for the possibility of diffractive proton excitation in
intermediate states, such as $p\to N^{*}$. However, it is possible incorporate the $s$-channel unitarity with elastic and a low mass
intermediate state $N^{*}$ by using a two-channel eikonal approach. The Good-Walker formalism \cite{goodWalker,paolo001,victor001,kmr001,kmr002} provides an elegant and
convenient form to incorporate $p\to N^{*}$ diffractive dissociation. In this approach, we introduce diffractive eigenstates
$|\phi_{i}\rangle$ that diagonalize the interaction matrix \textsf{T} (where $\textsf{S}=\textsf{1}+i\textsf{T}$). As a result, the
incoming hadron wave functions $| h \rangle$ (in our case
the ``beam'' and ``target'' proton wave functions) can be written as superpositions of these diffractive eigenstates, namely,
\begin{eqnarray}
| h \rangle_{beam} = \sum a_{i}|\phi_{i}\rangle, \hspace{0.8truecm} | h \rangle_{target} = \sum a_{k}|\phi_{k}\rangle .
\end{eqnarray}
Since we need at least two diffractive eigenstates, in a two-channel eikonal model, we have $i,k=1,2$. The extension to $n$-channel
eikonal models is straightforward; however, it is well known that a two-channel model is sufficient to capture the single- or
double-diffractive dissociation behavior very accurately \cite{kmr001,kmr002,kmr003,kmr004,kmr005,gotsman2,gotsman3}. In this paper, we
adopt a two-channel eikonal model in which the Pomeron couplings to the two diffractive eigenstates $k$ are
\begin{eqnarray}
\beta_{\Bbb P,k}(t)=(1\pm \gamma)\beta_{\Bbb P}(t) ,
\end{eqnarray}
i.e., the eigenvalues of the two-channel vertex are $1\pm \gamma$, where $\gamma \simeq 0.55$ \cite{kmr002,kmr003}. This value is in
accordance with $p \to N^{*}$ dissociation observed at CERN-ISR energies, more specifically, it is the value required in order to obtain
the experimental value of the cross section for low-mass diffraction measured at $\sqrt{s}=31$ GeV, namely,
$\sigma_{SD}^{lowM} \simeq 2$ mb.

Since each amplitude has two vertices, the forward observables are controlled by an elastic scattering amplitude with three different
exponents,
\begin{eqnarray}
{\cal A}_{eik}(s,t) &=& is \int^{\infty}_{0} b\, db\, J_{0}(bq) \left[ 1 -\frac{1}{4}\, e^{i(1+\gamma)^{2}\chi (s,b)} \right. \nonumber \\
& & \left. -\frac{1}{2}\, e^{i(1+\gamma^{2})\chi (s,b)} - \frac{1}{4}\, e^{i(1-\gamma)^{2}\chi (s,b)} \right] .
\label{doubeik001}
\end{eqnarray}
In the computation of the eikonal functions employed for calculating the amplitude above, again the input Born-level
amplitudes are the ones related to BI, BII, and BIII models. These two-channel eikonal models are referred to, respectively, as TI, TII,
and TIII models.

\section{Correlated Experimental Systematics}

In our analyses, we carry out global fits to forward $pp$ and $\bar{p}p$ scattering data above $\sqrt{s} = 10$ GeV and to
elastic $pp$ differential scattering cross section data at LHC energies. Specifically, we fit to the total cross section
$\sigma_{tot}^{pp,\bar{p}p}$, the ratio of the real to imaginary part of the forward scattering amplitude $\rho^{pp,\bar{p}p}$ and to the
elastic differential cross section $d\sigma^{pp}/dt$ at $\sqrt{s}=7$, 8, and 13 TeV with $|t|\leq 0.1$ GeV$^{2}$ (this range for
$|t|$ is enough for an appropriate evaluation of $\alpha^{\prime}_{\Bbb P}$). We use data sets compiled and analyzed by the Particle Data
Group (PDG) \cite{PDG} as well as the recent data at LHC from the TOTEM Collaboration
\cite{TOTEM008,TOTEM001,TOTEM002,TOTEM003,TOTEM004,TOTEM005,TOTEM007,TOTEM006,TOTEM009}, 
with the statistic and systematic errors added in quadrature. 

The PDG database is currently the standard data source used by most research papers in the field.
In order to ensure the consistency between TOTEM and PDG information, 
we must adopt common criteria for the
selection and treatment of data. 
Following the PDG discussion on treatment of errors, we see that correlated errors are indeed treated explicitly in the presence of
results of the form $A_{i}\pm \sigma_{i}\pm \Delta$ that have same systematic errors $\Delta$. As usual, it is possible to
average the
$A_{i}\pm \sigma_{i}$ and then combine in quadrature the resulting statistical error with the respective $\Delta$. However, the same
result can be obtained by averaging $A_{i}\pm \left( \sigma_{i}^{2} + \Delta_{i}^{2} \right)^{1/2}$, where $\Delta_{i}$ are modified
systematic errors given by
\begin{eqnarray}
\Delta_{i} =\sigma_{i} \Delta \left( \sum \frac{1}{\sigma_{j}^{2}}\right)^{1/2}.
\label{exp1pdg}
\end{eqnarray}
As pointed out by PDG, this alternative procedure has the advantage that each measurement may be treated as independent and, as a
consequence may be averaged in the usual way with other data.  
We will therefore adopt this procedure in preparing the data set to be fitted.

Let us start with the original TOTEM data, 
which include the first and second measurements of the total proton-proton ($pp$) cross section at $\sqrt{s}=7$ TeV,
$\sigma_{tot}^{pp}=98.3\pm 2.8$ mb \cite{TOTEM001}, and $\sigma_{tot}^{pp}=98.6\pm2.2$ mb \cite{TOTEM002} (both using the optical theorem
together with the luminosity provided by the CMS \cite{CMSlum}); the luminosity-independent measurement at $\sqrt{s}=7$ TeV,
$\sigma_{tot}^{pp}=98.0\pm2.5$ mb \cite{TOTEM003}; the $\rho$-independent measurements at $\sqrt{s}=7$ TeV of $\sigma_{tot}^{pp}$ and
$\rho$ parameter, $\sigma_{tot}^{pp}=99.1\pm 4.3$ mb and $\rho^{pp}=0.145\pm0.091$ \cite{TOTEM003}; the luminosity-independent
measurement at $\sqrt{s}=8$ TeV, $\sigma_{tot}^{pp}=101.7\pm2.9$ mb \cite{TOTEM004}; the measurements in the Coulomb-nuclear
interference region at $\sqrt{s}=8$ TeV of $\sigma_{tot}^{pp}$ and $\rho$ parameter, $\sigma_{tot}^{pp}=102.9\pm2.3$ mb and
$\sigma_{tot}^{pp}=103.0\pm2.3$ mb (for central and peripheral phase formulations, respectively), and $\rho^{pp}=0.12\pm0.03$
\cite{TOTEM005}; the total cross sections at $\sqrt{s}=8$ TeV, $\sigma_{tot}^{pp}=101.5\pm2.1$ mb and $\sigma_{tot}^{pp}=101.9\pm2.1$ mb,
obtained from extrapolations of the differential cross section to $t=0$ (for quadratic and cubic polynomials in the exponent,
respectively) \cite{TOTEM007}; the luminosity-independent measurements at $\sqrt{s}=13$ TeV, $\sigma_{tot}^{pp}=110.6\pm3.4$ mb, and
$\sigma_{tot}^{pp}=109.5\pm3.4$ mb \cite{TOTEM010}; the first extraction of the 
$\rho$ parameter at $\sqrt{s}=13$ TeV (exploiting the Coulomb-nuclear interference), $\rho^{pp}=0.09\pm0.01$ and $\rho^{pp}=0.10\pm0.01$,
and the associated total cross section $\sigma_{tot}^{pp}=110.3\pm3.5$ mb, obtained from the Coulomb normalization technique
\cite{TOTEM008}; the elastic differential cross section in the intervals $0.377 \le |t| \le 2.443$ GeV$^{2}$ 
\cite{TOTEM006} and $0.00515 \le |t| \le 0.235$ GeV$^{2}$ \cite{TOTEM003} at $\sqrt{s}=7$ TeV,
in the interval $6\times 10^{-4} \le |t| \le 0.2$ GeV$^{2}$ at $\sqrt{s}=8$ TeV \cite{TOTEM005}, and in the interval 
$0.0384 \le |t| \le 3.829$ GeV$^{2}$ at $\sqrt{s}=13$ TeV \cite{TOTEM009}.

From the $\sigma_{tot}^{pp}$ data at $\sqrt{s}=7$ TeV, there is already a hint of some kind of correlation among the measurements.
The four measurements taken by the TOTEM
group were obtained using the same beam optics configuration, namely, $\beta^{*} = 90$ m.
The optics with this betatron value is
very insensitive to variations of the machine parameters and led to very low systematic uncertainties on horizontal and vertical
scattering angles \cite{TOTEM001}.
Nevertheless, the value $\sigma_{tot}^{pp}=98.3\pm 2.8$ mb was obtained from the lower-luminosity run in June 2011,
whereas the remaining three values, namely, $\sigma_{tot}^{pp}=98.6\pm 2.2$ mb, $\sigma_{tot}^{pp}=98.0\pm 2.5$ mb, and 
$\sigma_{tot}^{pp}=99.1\pm 4.3$ mb, were
obtained in a dedicated run in October 2011. 
The October run has resulted in an improved measurement of the $t$-distribution with higher
statistics. Thus, although there is no correlation between the June and October measurements, the correlation among the last three
values of $\sigma_{tot}^{pp}$ became clear: they were obtained {\it from the same data}, recorded in the same run. Precisely, this type
of correlation also occurs in some $\sigma_{tot}^{pp}$ and $\rho^{pp}$ TOTEM data at $\sqrt{s}=8$ and 13 TeV.

The correlation issue also appears from reanalyses of part of the TOTEM data. In one of these works \cite{panchsriva01},
it is observed that a zero in the real part of the nuclear amplitude lies in the CNI region, leading to a positive amplitude at $-t=0$.
It has been proved some time ago \cite{martin001} that in the limit $s\to \infty$, if the total cross section tends to infinity and
the differential elastic cross
section tends to zero as $-t \gg 1$, the real part of the even amplitude must change sign near $-t=0$. Since these assumptions
correspond to the experimentally observed behavior of the cross sections at high energies, the $\rho$ value obtained by TOTEM assuming
a constant real part of the nuclear amplitude near $-t=0$ might be underestimated. Thus, in \cite{panchsriva01}, by analyzing the
complete TOTEM elastic differential cross section in the CNI region at $\sqrt{s} = $ 8 and 13 TeV, it is shown that two modified Barger
and Phillips (BP) amplitudes \cite{Barger001} appear to describe quite adequately the CNI data. The $\rho$ values obtained from these amplitudes
are
$\rho^{pp} = 0.135$ and $\rho^{pp} = 0.137$ at $\sqrt{s} = 8$ TeV as well as
$\rho^{pp} = 0.133$ and $\rho^{pp} = 0.134$ at $\sqrt{s} = 13$ TeV. Of course, these results are correlated to the TOTEM ones
since the same data set is used in their determinations.
The modified BP amplitudes also lead to different $\sigma_{tot}$
values than those obtained by TOTEM: $\sigma_{tot}^{pp}=102.7$ mb and
$\sigma_{tot}^{pp}=101.6$ mb at $\sqrt{s} = 8$ TeV, and $\sigma_{tot}^{pp}=112.9$ mb and
$\sigma_{tot}^{pp}=111.8$ mb at $\sqrt{s} = 13$ TeV. Once again, we have correlation among these and the TOTEM cross sections
result at $\sqrt{s} = 8$ and 13 TeV. In other reanalysis of the TOTEM data \cite{ezhelapetr01}, using a modified formula for the CNI
term, the values $\rho = 0.123\pm 0.010$ and $\sigma_{tot}^{pp}= 111.4 \pm 1.8$ mb are obtained at $\sqrt{s} = 13$ TeV. 

Since all results for $\rho^{pp}$ and $\sigma_{tot}^{pp}$ in \cite{panchsriva01} are obtained from the same data source
used by TOTEM, we would expect their associated errors to be the same as those of the TOTEM papers. Thus, in order to implement a
practical procedure for introducing correlation information in our fits, we will consider that the total and
systematic errors associated with the results in \cite{panchsriva01} are the same as those presented by TOTEM. This assumption makes
the PDG procedure be consistent, in good approximation, to the rule 
adopted by experimentalists in
accounting for correlation systematics, namely, to the product $\sqrt{n}\, \delta_{k}$, $k = 1, ..., n$, where $n$ is the number of correlated results
and $\delta_{k} = \sqrt{\sigma_{k}^{2} + \Delta^{2}}$. 
Thus, for practical reasons, we have adopted this last procedure for the selection and
treatment of correlated data. The set of correlated data used in our global fitting analyses is summarized in Table~\ref{tab001}.
In Fig. 1, we show these data with their corresponding uncertainties. The inner error bars are the original (published) uncertainties and
the outer error bars are the uncertainties after introducing correlation information.

\section{Results and Conclusions}

In all the fits presented in this paper, we use a $\chi^{2}$ fitting procedure,
where the value of $\chi^{2}_{min}$ is distributed as a $\chi^{2}$ distribution with N degrees of freedom (d.o.f). The fits to the
experimental data sets are performed adopting an interval $\chi^{2}-\chi^{2}_{min}$ corresponding, in the case of normal errors, to the
projection of the $\chi^{2}$ hypersurface containing 90\% of probability. This corresponds to $\chi^{2}-\chi^{2}_{min}=12.02$ and $13.36$
in the case of seven and eight free parameters, respectively.
Following the philosophy of using the minimum number of free parameters, in the following analyses, the slopes of the secondary-Reggeon
linear trajectories, $\alpha^{\prime}_{+}$ and $\alpha^{\prime}_{-}$, are fixed at 0.9 GeV$^{-1}$. These values are in agreement with those
usually obtained in Chew-Frautschi plots. Also, the slopes associated with the form factors of the secondary Reggeons are fixed at
$r_{+} = r_{-} = 4.0$ GeV$^{-2}$. 
These parameters have very little statistical correlation with the Pomeron
parameters, and their fixed values are consistent with those obtained in previous studies \cite{goulianos001,kmr002,kmr003}. We also
fix the scale of the pion-Pomeron vertex at $a_{1}=m_{\rho}^{2}=(0.776$ GeV)$^{2}$ \cite{kmr004}.

In the case of Born-level amplitudes, the
values of the Regge parameters determined by global fits to $pp$ and $\bar{p}p$ data are
listed in Table~\ref{tab002}. Notice that in the case of BI, BII, and BI+$\Bbb P \Bbb P$ models we fixed the parameter $r_{\Bbb P}$ at 5.5 GeV$^{-2}$,
which corresponds to the slope of the electromagnetic proton form factor. As discussed in Ref. \cite{kmr001}, it is the natural choice
for the computation of double-diffractive central Higgs production via $WW$ fusion (since the $W$ boson is radiated from a quark, like
the photon). Moreover, our analyses show that at this $r_{\Bbb P}$ value the Pomeron is described by trajectories with
$\alpha^{\prime}_{\Bbb P} \simeq 0.25$  GeV$^{-2}$ (see Table~\ref{tab002}). Interestingly enough, these values for $\alpha^{\prime}_{\Bbb P}$ are
consistent with the ones recently obtained from holographic QCD models \cite{alfonso001}. Furthermore, if we perform the global fits at
another value of $r_{\Bbb P}$,
say 4.0 GeV$^{-2}$ (which is not atypical \cite{kmr001}), we obtain the values $\alpha^{\prime}_{\Bbb P} = $ 0.3346$\pm$0.0085 GeV$^{-2}$,
0.3339$\pm$0.0085 GeV$^{-2}$, and $\alpha^{\prime}_{\Bbb P} = $ 0.3346$\pm$0.0090 GeV$^{-2}$ in the case of BI, BII, and BI+$\Bbb P \Bbb P$
models, respectively, while the remaining free parameters follow with the same values. The parameters obtained in the analyses with 
BI, BII, and BI+$\Bbb P \Bbb P$ models are very close to each other and the description of the data resulted in
substantially the same curves, shown in Fig. 2 (solid curves). The dashed curves in the same figure are for the BIII model. Figures 2-4
have the same layout: the part (a) shows the $\sigma_{tot}^{pp,\bar{p}p}$ accelerator data, the part (b) extends the range in
$\sqrt{s}$ of the part (a), the part (c) shows the
$\rho^{pp,\bar{p}p}$
data, and the part  (d) extends the range in $\sqrt{s}$ of the part (c). For comparison purposes, we have included in part (b) of these figures
estimates of $\sigma_{tot}^{pp}$ obtained from cosmic ray experiments, namely, the AUGER experimental result at $\sqrt{s} = 57$ TeV \cite{auger} and the Telescope Array result at $\sqrt{s} = 95$ TeV \cite{TA}.

The preceding results using Born-level amplitudes have demonstrated that it is possible a
good description of forward data up to LHC energy by using a constant value of $\epsilon$; even so, from the Table~\ref{tab002}, we see that the ratio of two-Pomeron to one-Pomeron exchange couplings is not so small,
\begin{eqnarray}
R \equiv \frac{\left| {\Bbb P\Bbb P}_{\textnormal{coupling}} \right|}{{\Bbb P}_{\textnormal{coupling}}} = \frac{\beta^{2}_{\Bbb P\Bbb P}(0)}{\beta^{2}_{\Bbb P}(0)} \simeq
0.6 ,
\end{eqnarray}
which suggests the violation of unitarity at the presently available energies. Therefore, contrary to the expectation that the violation of unitarity would occur only far above the LHC energies, the value of $R$ indicates that the unitarization breaking is a current problem. Most importantly, this result indicates that multi-Pomeron exchanges must be included in order to restore unitarity. Multiple exchanges can somehow be taken into account in an appropriate unitarization scheme, such as the eikonalization one,
as already discussed.

The values of the Regge parameters obtained using one-channel eikonal models are listed in Table~\ref{tab003}. We see that all eikonalized fits to the data prefer very small values of $\alpha^{\prime}_{I\!\!P}$, in the case of OI, OII, and OIII models these values are, respectively, $\alpha^{\prime}_{I\!\!P} = 0.066\pm 0.012$, $0.039\pm 0.012$, and $0.052\pm 0.029$ GeV$^{-2}$. Thus, after the eikonalization, the Pomeron looks similar to a fixed pole at $\alpha_{I\!\!P}(0) = 1$. We can see that the effect of the pion-loop insertions further decreases the values
of $\alpha^{\prime}_{I\!\!P}$: the result for OII (OIII) model is about a factor 1.7 (1.3) lower than the OI one. We also note a substantial increase of the Pomeron intercept in relation to Born-level models: the $\epsilon$ parameters are increased by about 34\% for OI and OII models and by about 38\% for the OIII model.
The curves of $\sigma_{tot}(s)$ and $\rho (s)$ for the one-channel models, compared with the experimental
data, are shown in Fig. 3.

The same significant increase of the Pomeron intercept in relation to Born-level models is observed in the two-channel eikonal models, as shown in Table~\ref{tab004}. In the case of TI, TII, and TIII models,
the increase is even greater, with the $\epsilon$ parameters being about 61\%, 60\%, and 53\% higher, respectively, when compared to the same parameters in models with Born-level amplitudes. Again, we see that the eikonal fits to the data prefer very small values of $\alpha^{\prime}_{I\!\!P}$, but now with values of $\alpha^{\prime}_{I\!\!P}$ closer to each other. In the case of TI, TII, and TIII models, these values are, respectively, $\alpha^{\prime}_{I\!\!P} = 0.046\pm 0.012$, $0.048\pm 0.012$, and $0.050\pm 0.029$ GeV$^{-2}$. The results for $\sigma_{tot}(s)$ and $\rho (s)$, using two-channel eikonal amplitudes, are shown in Fig. 4.

For both one- and two-channel eikonal models, the values of the parameter $r_{\Bbb P}$ are insensitive to changes in the form of the Pomeron trajectory, as they are of the same order. 
In the case of one-channel models, the change from a linear to a nonlinear Pomeron trajectory leads to an increase of $r_{\Bbb P}$ of only 7\%, while in two-channel models the same change leaves the value of $r_{\Bbb P}$ practically unchanged. In Table~\ref{tab005}, we show the predictions on the high energy total cross section and $\rho$ parameter, made using the BIII, OIII, and TIII models. A comparison among the cross sections and $\rho$ parameters at high energies, from the models BIII (solid), OIII (dashed), and TIII (dotted), is shown in Fig. 5.

In Fig. 6, we show the description of the elastic $pp$ differential cross section data for all models discussed so far. We see that in all cases
the data in the range $|t|_{min} \leq |t| \leq 0.1$ GeV$^{2}$ are well described, where we have adopted $|t|_{min} \sim 10|t|_{int}$, since in the region $|t| \gg |t|_{int}$ the nuclear scattering dominates. The value of $|t|_{int}$ (where the interference between the Coulomb and hadronic amplitudes is of maximum significance) can be simply determined from the practical relation $|t|_{int} = 0.071/\sigma_{tot}$ \cite{blockcahn}. In Figs. 6(a)-6(c), we show the curves of $d\sigma^{pp}/dt$ for the Born level, one- and two-channel eikonal models, respectively. We can see that it is not possible to distinguish between different models of the Pomeron since the curves representing models of types I (solid), II (dashed), and III (dotted) fall on very nearly the same curve as the model of type I. A comparison among the differential cross sections from the models BIII (solid), OIII (dashed), and TIII (dotted) is shown in Fig. 6(d).

In order to see more clearly the effects of the correlation information on the data sets, we compare, in Fig. 2(b), the total $pp$ cross section data of TOTEM with the ATLAS data. The ATLAS results include the luminosity-dependent measurements at $\sqrt{s}=7$ TeV, $\sigma_{tot}^{pp}=95.35\pm 1.36$ \cite{atlas001} [open circle in Fig. 2(b)], and $\sqrt{s}=8$ TeV, $\sigma_{tot}^{pp}=96.07\pm 0.92$ \cite{atlas002} [open square in Fig. 2(b)] . These measurements rely on the optical theorem. The luminosity, necessary to normalize the elastic cross section, is determined from LHC beam parameters using van der Meer scans \cite{atlas001a}. Before the introduction of correlation information, we recognize some tension between the TOTEM and ATLAS measurements. For example, if we compare the ATLAS result for $\sigma_{tot}^{pp}$ at $\sqrt{s}=7$ TeV with, say, the value $\sigma_{tot}^{pp}=98.6\pm 2.2$
measured by TOTEM at the same energy, the difference between the values corresponds to 1.5 $\sigma$. In the same way, if we compare the
ATLAS result for $\sigma_{tot}^{pp}$ at $\sqrt{s}=8$ TeV with the highest value obtained by TOTEM in the same energy, $\sigma_{tot}^{pp}=103.0\pm 2.3$, the difference goes up to 3 $\sigma$. However, after the inclusion of the correlated systematics that allows treating each result as independent, the difference between the ATLAS results and the average values of the TOTEM measurements and uncertainties at 7 and 8 TeV drops to 0.7 and 1.6 $\sigma$, respectively. We can therefore ameliorate the tension between the TOTEM and ATLAS measurements by introducing correlation information.

In summary, the paper is devoted to a detailed study of the soft Pomeron. More precisely, we evaluate the
relative plausibility of different combinations of vertices and trajectories of the soft Pomeron in the light of the recent LHC
data, taking into account the existent systematic correlation among them. The methods used in this paper for the examination of the Pomeron properties can also be applied to obtain constraints on the Odderon parameters. Work in this direction is in progress.

\section*{Acknowledgment}

This research was partially supported by the Agencia Nacional de Investigaci\'on e Innovaci\'on under the project ANII-FCE-126412 and by the Conselho Nacional de Desenvolvimento Cient\'{\i}fico e Tecnol\'ogico under Grants No. 141496/2015-0 and No. 155628/2018-6.

\newpage

\begin{table*}[h!]
\centering
\caption{LHC data used in our global fitting analyses. In the case of correlated data, uncertainties are multiplied by a factor
$f_{n}\equiv \sqrt{n}$,
where $n$ is the number of correlated quantities.}
\begin{tabular}{c@{\quad}|c@{\quad}|c@{\quad}|c@{\quad}}
\hline \hline 
& & & \\[-0.3cm]
$\sqrt{s}$ (TeV) & $\sigma_{\text{tot}}$ (mb) & $\rho$ \\ 
\hline 
& & & \\[-0.3cm]
13 & $110.6 \pm (3.4 \times f_{5})$ \cite{TOTEM010}     & $0.100\pm (0.010 \times f_{4})$ \cite{TOTEM008} \\
   & $109.5 \pm (3.4 \times f_{5})$ \cite{TOTEM010}     & $0.133\pm (0.010 \times f_{4})$ \cite{panchsriva01} \\
   & $111.8 \pm (3.4 \times f_{5})$ \cite{panchsriva01} & $0.134\pm (0.010 \times f_{4})$ \cite{panchsriva01} \\
   & $112.9 \pm (3.4 \times f_{5})$ \cite{panchsriva01} & $0.123\pm (0.010 \times f_{4})$ \cite{ezhelapetr01} \\
   & $111.4 \pm (1.8 \times f_{5})$ \cite{ezhelapetr01} & $0.090\pm (0.010 \times f_{3})$ \cite{TOTEM008} \\
   & $110.3 \pm 3.5$ \cite{TOTEM008} & $0.133\pm (0.010 \times f_{3})$ \cite{panchsriva01}  \\
   &  & $0.134\pm (0.010 \times f_{3})$ \cite{panchsriva01}  \\
\hline
& & & \\[-0.3cm]
8  & $102.9\pm (2.3 \times f_{4})$ \cite{TOTEM005} & $0.120\pm (0.030 \times f_{3})$ \cite{TOTEM005} \\
  & $103.0\pm (2.3 \times f_{4})$ \cite{TOTEM005} & $0.137\pm (0.030 \times f_{3})$ \cite{panchsriva01} \\ 
  & $101.6\pm (2.3 \times f_{4})$ \cite{panchsriva01} & $0.135\pm (0.030 \times f_{3})$ \cite{panchsriva01} \\  
  & $102.7\pm (2.3 \times f_{4})$ \cite{panchsriva01} &  \\ 
  & $101.5\pm (2.1 \times f_{2})$ \cite{TOTEM007} & \\  
  & $101.9\pm (2.1 \times f_{2})$ \cite{TOTEM007} & \\ 
  & $101.7\pm 2.9$ \cite{TOTEM004} &  \\
\hline
& & & \\[-0.3cm]
7  & $99.1\pm (4.3 \times f_{3})$ \cite{TOTEM003} & $0.145\pm 0.091$ \cite{TOTEM003} \\
  & $98.0\pm (2.5 \times f_{3})$ \cite{TOTEM003} & \\
  & $98.6\pm (2.2 \times f_{3})$ \cite{TOTEM002} &  \\
  & $98.3\pm 2.8 $ \cite{TOTEM001} &  \\
\hline \hline 
\end{tabular} 
\label{tab001}
\end{table*}

\begin{table*}
\centering
\caption{TOTEM. The values of the Pomeron and secondary Reggeon parameters obtained in global fits to the $\sigma_{tot}^{pp,\bar{p}p}$,
$\rho^{pp,\bar{p}p}$, and $d\sigma^{pp,\bar{p}p}/dt$ data using Born-level amplitudes. The parameters
$\alpha^{\prime}_{+}$, $\alpha^{\prime}_{-}$, $r_{+}$, $r_{-}$, and $a_{1}$ are fixed.}
\begin{ruledtabular}
\begin{tabular}{ccccc}
 & \multicolumn{4}{c}{Born-level amplitudes}  \\
\cline{2-5}
 & BI & BII & BIII & BI + $\Bbb P\Bbb P$  \\
\hline
$\epsilon$ & 0.0942$\pm$0.0030 & 0.0943$\pm$0.0030 & 0.0949$\pm$0.0032 & 0.101$\pm$0.012 \\
$\alpha^{\prime}_{I\!\!P}$ (GeV$^{-2}$) & 0.249$\pm$0.014 & 0.248$\pm$0.014 & 0.3210$\pm$0.0016 & 0.175$\pm$0.084 \\
$\beta_{\Bbb P}(0)$ (GeV$^{-1}$) & 1.956$\pm$0.048 & 1.955$\pm$0.048 & 1.948$\pm$0.052 & 2.034$\pm$0.087 \\
$r_{\Bbb P}$ (GeV$^{-2}$) & 5.5 (fixed) & 5.5 (fixed) &  $\cdots$  & 5.5 (fixed) \\
$\eta_{+}$ & 0.338$\pm$0.050 & 0.338$\pm$0.050 & 0.333$\pm$0.053 & 0.310$\pm$0.061 \\
$\beta_{+}(0)$ (GeV$^{-1}$) & 3.73$\pm$0.40 & 3.73$\pm$0.40 & 3.71$\pm$0.42 & 3.80$\pm$0.40 \\
$\eta_{-}$ & 0.529$\pm$0.085 & 0.529$\pm$0.085 & 0.528$\pm$0.089 & 0.524$\pm$0.089 \\
$\beta_{-}(0)$ (GeV$^{-1}$) & 2.90$\pm$0.52 & 2.90$\pm$0.52 & 2.90$\pm$0.54 & 2.88$\pm$0.53 \\
$a_{1}$ (GeV$^{2}$) & $\cdots$ & $m_{\rho}^{2}$ (fixed) & $m_{\rho}^{2}$ (fixed) & $\cdots$ \\
$a_{2}$ (GeV$^{2}$) & $\cdots$ & $\cdots$ & 1.79$\pm$0.44 & $\cdots$ \\
$\beta_{\Bbb P\Bbb P}(0)$ (GeV$^{-1}$) & $\cdots$ & $\cdots$ & $\cdots$ & 1.52$\pm$0.66 \\
\hline
$\chi^{2}/d.o.f$ & 0.65 & 0.65 & 0.65 & 0.63 \\
\end{tabular}
\end{ruledtabular}
\label{tab002}
\end{table*}

\begin{table*}
\centering
\caption{TOTEM. The values of the Pomeron and secondary Reggeon parameters obtained in global fits to the $\sigma_{tot}^{pp,\bar{p}p}$,
$\rho^{pp,\bar{p}p}$, and $d\sigma^{pp,\bar{p}p}/dt$ data using one-channel eikonalized amplitudes. The parameters
$\alpha^{\prime}_{+}$, $\alpha^{\prime}_{-}$, $r_{+}$, $r_{-}$, and $a_{1}$ are fixed.}
\begin{ruledtabular}
\begin{tabular}{cccc}
 & \multicolumn{3}{c}{Eikonalized amplitudes (one-channel eikonal)}  \\
\cline{2-4}
 & OI & OII & OIII  \\
\hline
$\epsilon$ & 0.1258$\pm$0.0014 & 0.1267$\pm$0.0049 & 0.1309$\pm$0.0087  \\
$\alpha^{\prime}_{I\!\!P}$ (GeV$^{-2}$) & 0.066$\pm$0.012 & 0.039$\pm$0.012 & 0.052$\pm$0.029  \\
$\beta_{\Bbb P}(0)$ (GeV$^{-1}$) & 1.811$\pm$0.010 & 1.795$\pm$0.035 & 1.77$\pm$0.12  \\
$r_{\Bbb P}$ (GeV$^{-2}$) & 6.53$\pm$0.23 & 7.0$\pm$5.2 & $\cdots$   \\
$\eta_{+}$ & 0.278$\pm$0.063 & 0.275$\pm$0.060 & 0.267$\pm$0.056  \\
$\beta_{+}(0)$ (GeV$^{-1}$) & 3.98$\pm$0.47 & 3.95$\pm$0.70 & 3.94$\pm$0.40  \\
$\eta_{-}$ & 0.534$\pm$0.087 & 0.531$\pm$0.088 & 0.531$\pm$0.087  \\
$\beta_{-}(0)$ (GeV$^{-1}$) & 3.41$\pm$0.64 & 3.39$\pm$0.71 & 3.40$\pm$0.62  \\
$a_{1}$ (GeV$^{2}$) & $\cdots$ & $m_{\rho}^{2}$ (fixed) & $m_{\rho}^{2}$ (fixed)  \\
$a_{2}$ (GeV$^{2}$) & $\cdots$ & $\cdots$ & 0.58$\pm$0.26  \\
\hline
$\chi^{2}/d.o.f$ & 0.66 & 0.66 & 0.66  \\
\end{tabular}
\end{ruledtabular}
\label{tab003}
\end{table*}

\begin{table*}
\centering
\caption{TOTEM. The values of the Pomeron and secondary Reggeon parameters obtained in global fits to the
$\sigma_{tot}^{pp,\bar{p}p}$, $\rho^{pp,\bar{p}p}$, and $d\sigma^{pp,\bar{p}p}/dt$ data using two-channel
eikonalized amplitudes.
The parameters $\alpha^{\prime}_{+}$, $\alpha^{\prime}_{-}$, $r_{+}$, $r_{-}$, and $a_{1}$ are fixed.}
\begin{ruledtabular}
\begin{tabular}{cccc}
 & \multicolumn{3}{c}{Eikonalized amplitudes (two-channel eikonal)}  \\
\cline{2-4}
 & TI & TII & TIII  \\
\hline
$\epsilon$ & 0.152$\pm$0.010 & 0.1513$\pm$0.0017 & 0.1544$\pm$0.0073  \\
$\alpha^{\prime}_{I\!\!P}$ (GeV$^{-2}$) & 0.0460$\pm$0.0085 & 0.048$\pm$0.023 & 0.0500$\pm$0.0053  \\
$\beta_{\Bbb P}(0)$ (GeV$^{-1}$) & 1.72$\pm$0.16 & 1.726$\pm$0.023 & 1.75$\pm$0.12  \\
$r_{\Bbb P}$ (GeV$^{-2}$) & 5.7$\pm$1.2 & 5.67$\pm$0.39 &  $\cdots$   \\
$\eta_{+}$ & 0.257$\pm$0.058 & 0.258$\pm$0.014 & 0.261$\pm$0.053  \\
$\beta_{+}(0)$ (GeV$^{-1}$) & 4.27$\pm$0.44 & 4.28$\pm$0.13 & 4.32$\pm$0.42  \\
$\eta_{-}$ & 0.537$\pm$0.086 & 0.535$\pm$0.073 & 0.538$\pm$0.086  \\
$\beta_{-}(0)$ (GeV$^{-1}$) & 3.80$\pm$0.69 & 3.79$\pm$0.59 & 3.83$\pm$0.69  \\
$a_{1}$ (GeV$^{2}$) & $\cdots$ & $m_{\rho}^{2}$ (fixed) & $m_{\rho}^{2}$ (fixed)  \\
$a_{2}$ (GeV$^{2}$) & $\cdots$ & $\cdots$ & 1.02$\pm$0.13  \\
\hline
$\chi^{2}/d.o.f$ & 0.71 & 0.71 & 0.70  \\
\end{tabular}
\end{ruledtabular}
\label{tab004}
\end{table*}

\begin{table*}
\centering
\caption{Predictions and uncertainties for the forward scattering quantities $\sigma_{tot}^{pp}$ and $\rho^{pp}$ using different Regge models.
The uncertainties are just estimates since the full covariance matrix was not employed.}
\begin{ruledtabular}
\begin{tabular}{ccccccc}
 & \multicolumn{2}{c}{BIII model} & \multicolumn{2}{c}{OIII model}& \multicolumn{2}{c}{TIII model} \\
\cline{2-3} \cline{4-5} \cline{6-7}
$\sqrt{s}$ (TeV) & $\sigma_{tot}$ (mb) & $\rho$ & $\sigma_{tot}$ (mb) & $\rho$ & $\sigma_{tot}$ (mb) & $\rho$ \\
\hline
$7.0$ & 98.7$\pm$2.7 & 0.149$\pm$0.008 & 99.2$\pm$2.7 & 0.135$\pm$0.008 & 98.8$\pm$2.7 & 0.134$\pm$0.008 \\
$8.0$ & 101.2$\pm$2.8 & 0.149$\pm$0.009 & 101.5$\pm$2.8 & 0.134$\pm$0.009 & 101.1$\pm$2.8 & 0.133$\pm$0.010 \\
$13.0$ & 110.9$\pm$3.3 & 0.150$\pm$0.009 & 110.2$\pm$3.3 &  0.130$\pm$0.009 & 109.6$\pm$3.3 & 0.129$\pm$0.012 \\
$57.0$ & 148$\pm$8 & 0.150$\pm$0.018 & 139$\pm$7 & 0.116$\pm$0.018 & 138.1$\pm$7 & 0.117$\pm$0.018 \\
$95.0$ & 162$\pm$12 & 0.150$\pm$0.026 & 150$\pm$11 & 0.112$\pm$0.026 & 149$\pm$11 & 0.113$\pm$0.026 \\
\end{tabular}
\end{ruledtabular}
\label{tab005}
\end{table*}

\begin{figure*}\label{fig006}
\begin{center}
\includegraphics[height=.70\textheight]{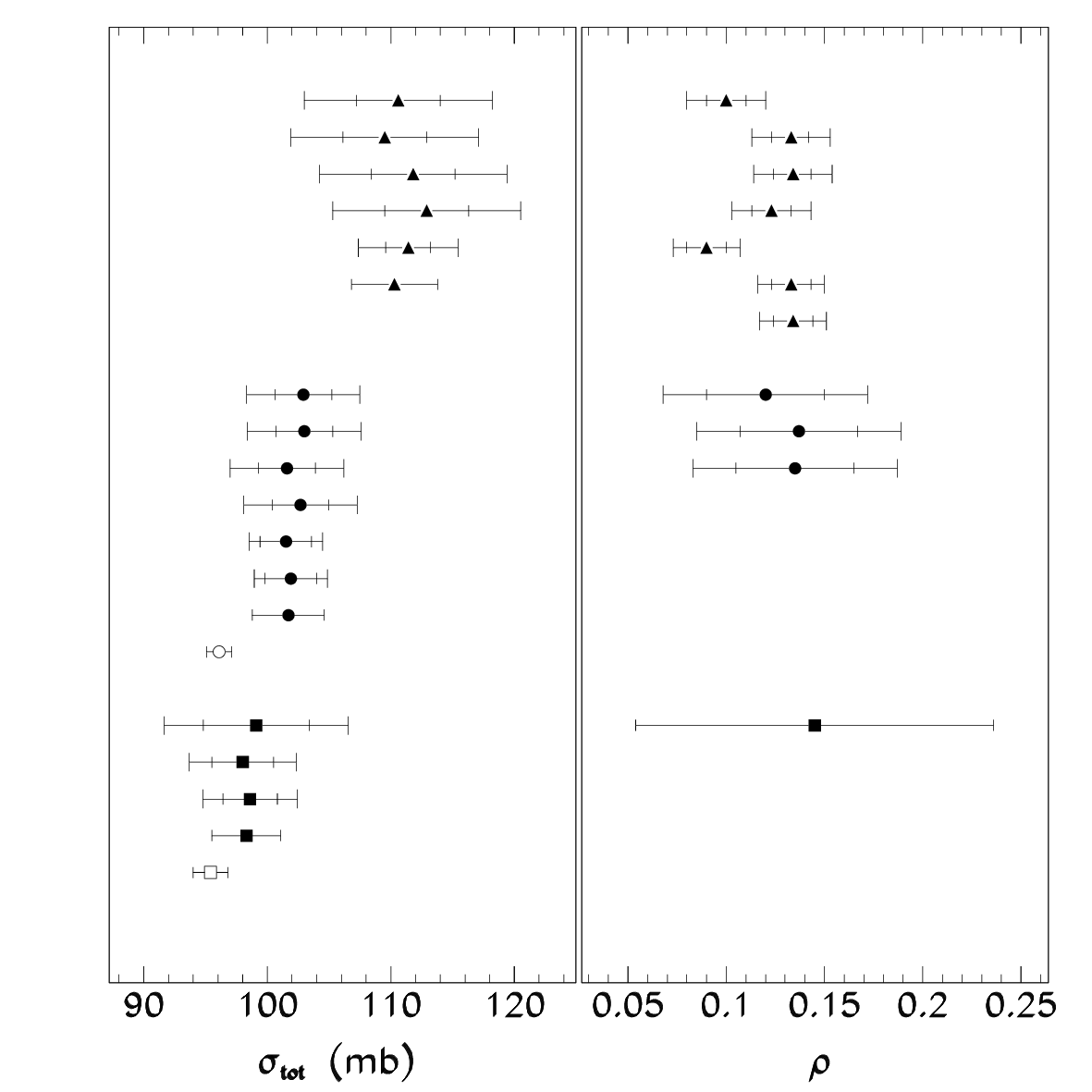}
\caption{Set of correlated data used in our global fitting analyses. The inner error bars are the original (published) uncertainties and
the outer error bars are the uncertainties after introducing correlation information. The squares, circles, and triangles correspond, respectively, to the energies of 7, 8, and 13 TeV. The solid symbols represent the TOTEM data, while the open ones represent the ATLAS results.}
\end{center}
\end{figure*}

\begin{figure*}\label{fig001}
\begin{center}
\includegraphics[height=.70\textheight]{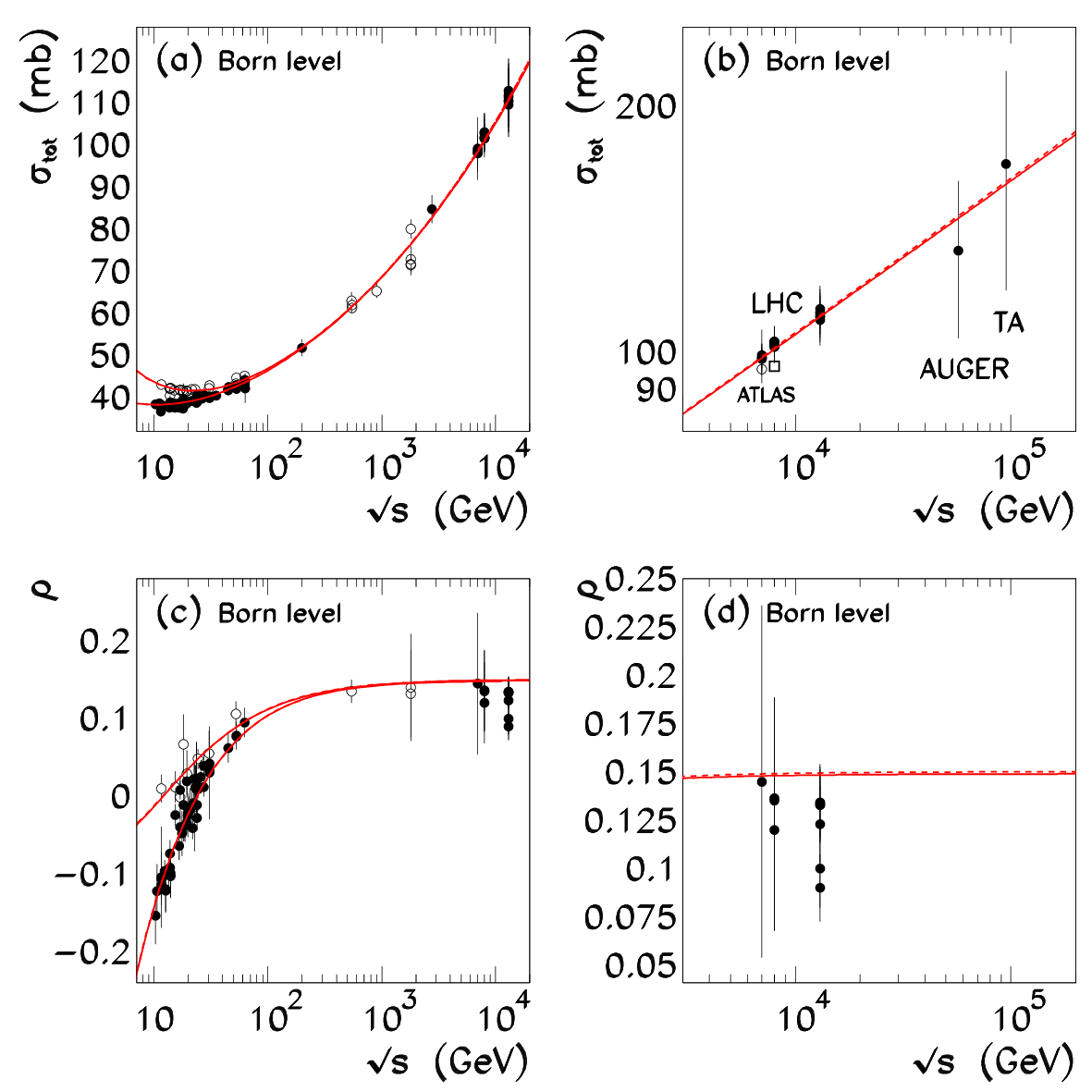}
\caption{Total cross section [(a) and (b)] and ratio of the real to imaginary part of the forward scattering amplitude
[(c) and (d)] for $pp$ ($\bullet$) and $\bar{p}p$ ($\circ$) channels. The solid line shows the results obtained using the BI or BII
or BI+$\Bbb P \Bbb P$ models, while the dashed line shows the results obtained using the BIII model. 
Also shown the predictions for cosmic ray energies.}
\end{center}
\end{figure*}

\begin{figure*}\label{fig002}
\begin{center}
\includegraphics[height=.70\textheight]{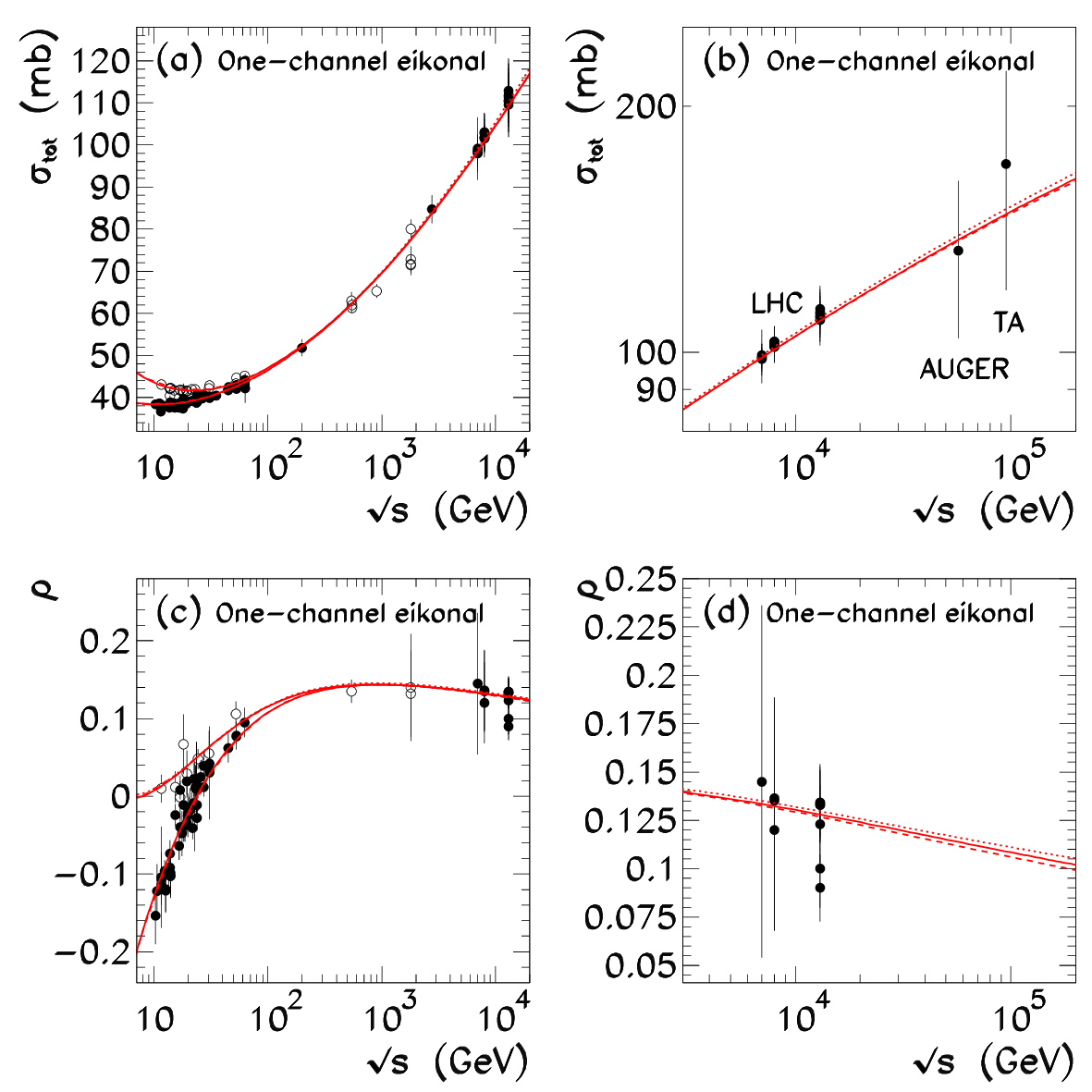}
\caption{Total cross section [(a) and (b)] and ratio of the real to imaginary part of the forward scattering amplitude
[(c) and (d)] for $pp$ ($\bullet$) and $\bar{p}p$ ($\circ$) channels. The solid, dashed, and dotted lines show the
results obtained using the OI, OII, and OIII models, respectively. 
Also shown the predictions for cosmic ray energies.}
\end{center}
\end{figure*}

\begin{figure*}\label{fig003}
\begin{center}
\includegraphics[height=.70\textheight]{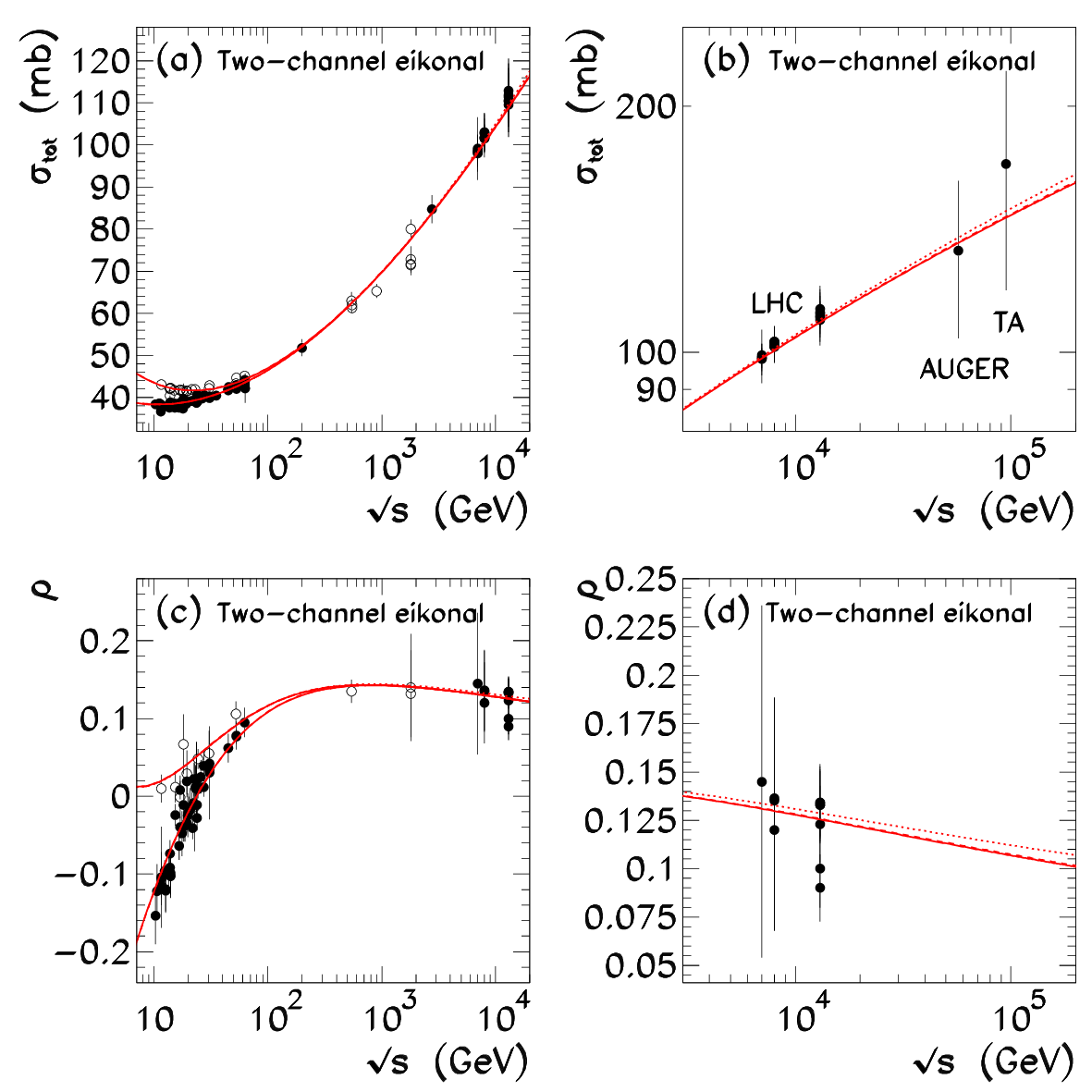}
\caption{Total cross section [(a) and (b)] and ratio of the real to imaginary part of the forward scattering amplitude
[(c) and (d)] for $pp$ ($\bullet$) and $\bar{p}p$ ($\circ$) channels. The solid, dashed, and dotted lines show the
results obtained using the TI, TII, and TIII models, respectively. 
Also shown the predictions for cosmic ray energies.}
\end{center}
\end{figure*}

\begin{figure*}\label{fig005}
\begin{center}
\includegraphics[height=.70\textheight]{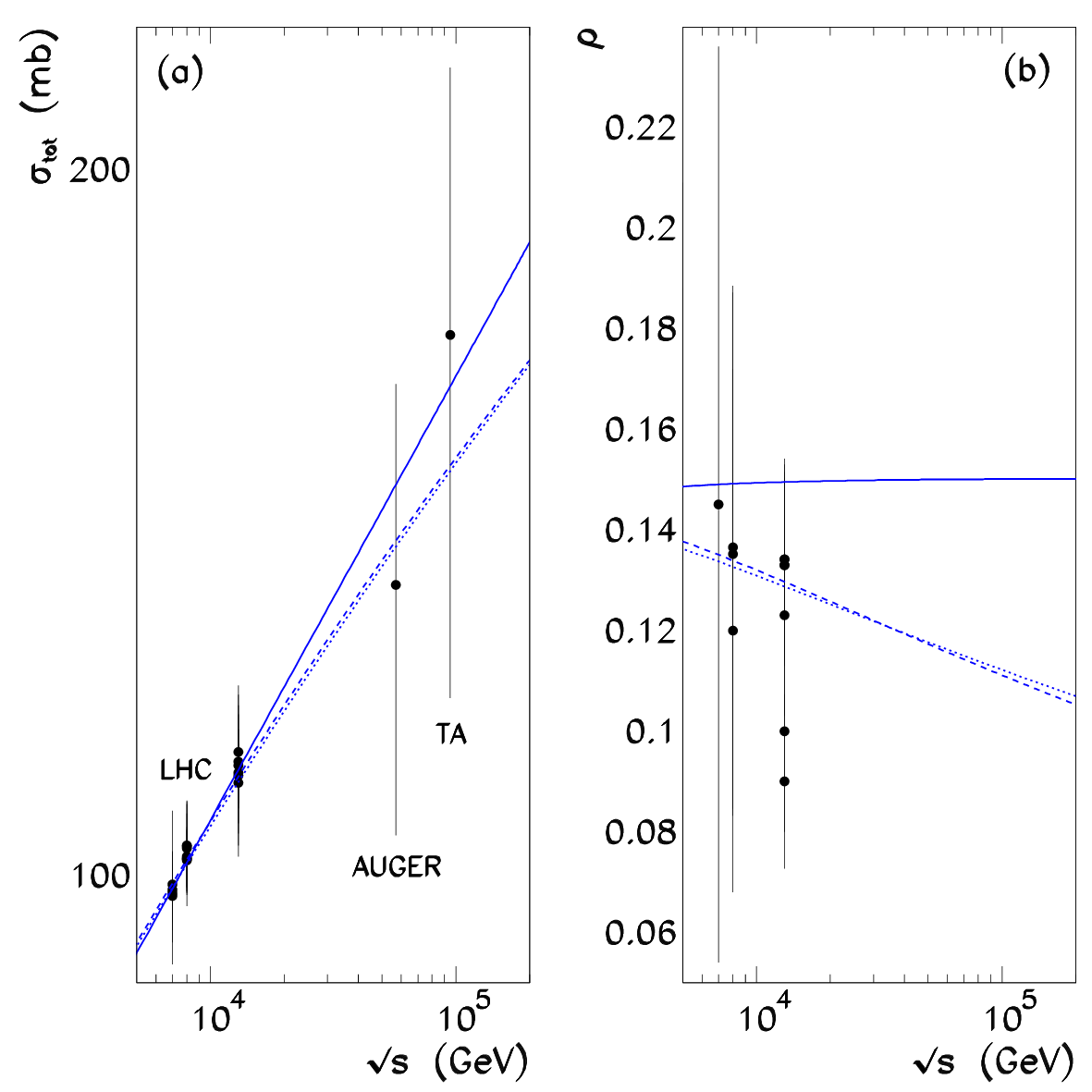}
\caption{Total cross section [part (a)] and ratio of the real to imaginary part of the forward scattering amplitude
[part (b)] for $pp$ ($\bullet$) channel. The solid, dashed, and dotted lines show the
results obtained using the BIII, OIII, and TIII models, respectively. 
Also shown the predictions for cosmic ray energies.}
\end{center}
\end{figure*}

\begin{figure*}\label{fig004}
\begin{center}
\includegraphics[height=.70\textheight]{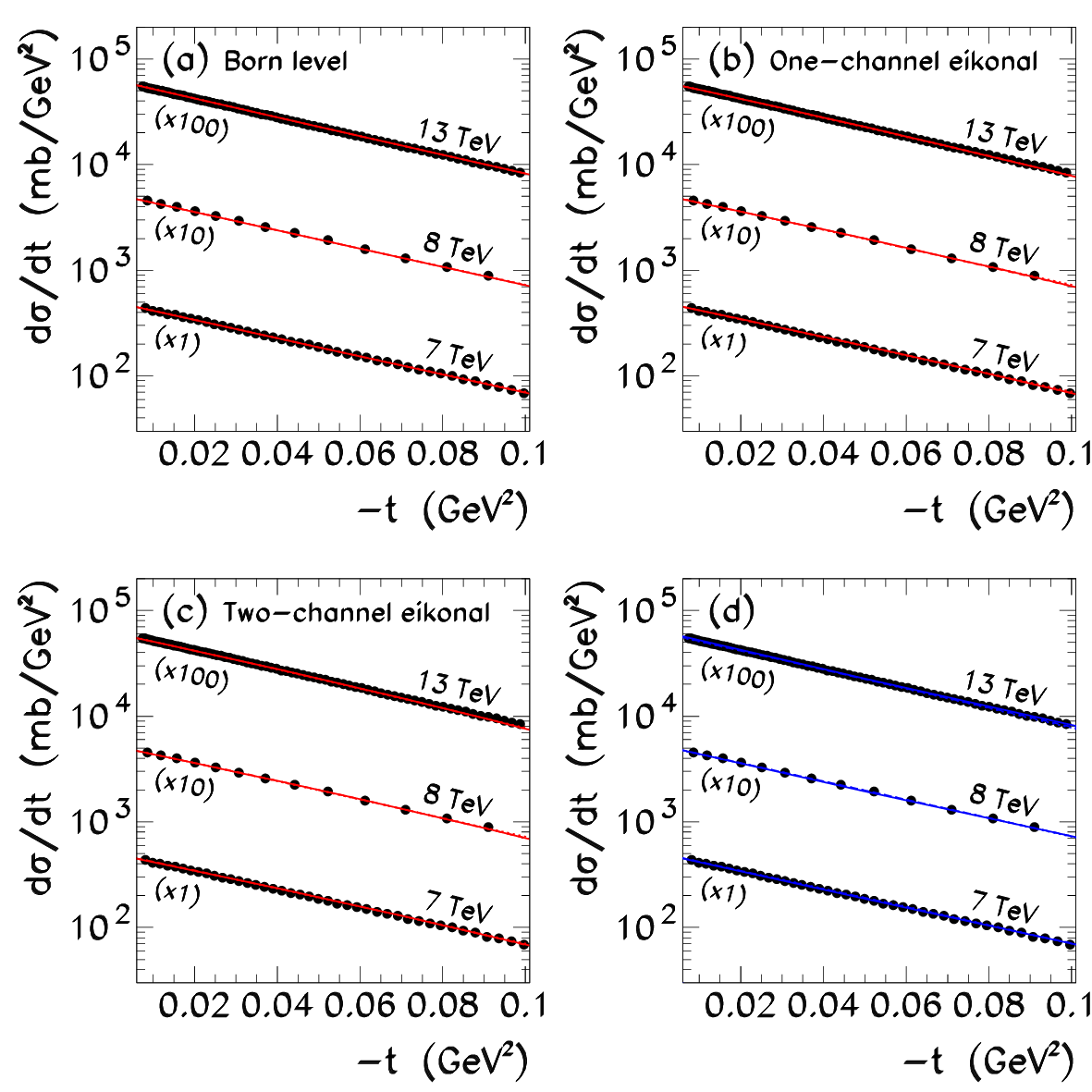}
\caption{The elastic differential cross section for $pp$ ($\bullet$) channel. In parts (a)-(c), we show the curves of $d\sigma^{pp}/dt$ for the Born level, one- and two-channel eikonal models, respectively. A comparison among the differential cross sections from the models BIII (solid), OIII (dashed), and TIII (dotted) is shown in part (d).}
\end{center}
\end{figure*}

\end{document}